# FROM CHIRAL ANOMALY TO TWO-FLUID HYDRODYNAMICS FOR ELECTRONIC VORTICES.

S. Brazovskii[a,c,1], N. Kirova[b,c,2]

[a]*LPTMS, UMR 8626, CNRS & Université Paris-sud, Univ. Paris-Saclay, Orsay, cedex, France*

[b]*LPS, UMR 8502, CNRS & Université Paris-sud, Univ. Paris-Saclay, Bat. 510, Orsay, cedex, France*

[c]*National University of Science and Technology "MISiS", 119049, Moscow, Russia.*

**Abstract.**

Many recent experiments addressed manifestations of electronic crystals, particularly the charge density waves, in nano-junctions, under electric field effect, at high magnetic fields, together with real space visualizations by STM and micro X-ray diffraction. This activity returns the interest to stationary or transient states with static and dynamic topologically nontrivial configurations: electronic vortices as dislocations, instantons as phase slip centers, and ensembles of microscopic solitons. Describing and modeling these states and processes calls for an efficient phenomenological theory which should take into account the degenerate order parameter, various kinds of normal carriers and the electric field. Here we notice that the commonly employed time-depend Ginzburg-Landau approach suffers with violation of the charge conservation law resulting in unphysical generation of particles which is particularly strong for nucleating or moving electronic vortices. We present a consistent theory which exploits the chiral transformations taking into account the principle contribution of the fermionic chiral anomaly to the effective action. The resulting equations clarify partitions of charges, currents and rigidity among subsystems of the condensate and normal carriers. On this basis we perform the numerical modeling of a spontaneously generated coherent sequence of phase slips - the space-time vortices - serving for the conversion among the injected normal current and the collective one.

---

[1] brazov@lptms.u-psud.fr
[2] kirova@lps.u-psud.fr  corresponding author

# 1. Introduction.

Embedded or transient topologically nontrivial configurations are common among symmetry broken ground states of electronic systems. Among them, the family of electronic crystals [1] (CDWs, SDWs, Wigner crystals and stripes arrays) demonstrates a universal ability of the collective motion giving rise to a rich complex of nonlinear and nonstationary behavior. Among various electronic crystals, the charge density waves (CDW) present the most popular and convenient object of studies [2,3] with only a small critical electric field necessary for creation of topologically nontrivial objects like solitons, vortices, dislocations, and phase slips. The role of dynamical electronic vorticity (phase slips as space-time vorticies, flows of dislocations) was inferred from experiments on space-resolved X-ray diffraction [4-8], in multi-contact [9] and mesa-junction [10] devices, and in coherent micro-diffraction [11, 12]. On the microscopic level, a breakthrough came from direct visualization of solitons by STM [13,14]. More evidences on solitons are coming from optical [15] and ac [16] measurements. On the theory side, considerations of vorticity in a lateral geometry [17-20] can be related to experiments on field-effect [21] or mesa [10] junctions. Random phase slips were inferred in the interpretation of dissipation peaks data collected by STM/AFM studies of a 2D CDW crystal [22]. Dynamical vorticity was modeled, within different schemes, for geometries of a 1D wire [23,24] and even of the real 2D mesa-junction [25-27]. Recently, the coherent sequences of phase slips were modeled and inferred from experiments on nano-junctions near the integer Hall state [28] where the CDW was set to the motion by the Hall voltage.

Kinetic theory of the sliding state was developed in several approaches, mostly limited to the linear response or a constant gap cases. An integrated approach with a review of previous achievements and contradictions can be found in [29,30] and with great details in [31]; the most systematic microscopic studies can be found in the review [32] and subsequent publications [33,34]. The goal of understanding and numerical modeling of the nonlinear phenomena requires for an efficient tool to go beyond the linear approach, particularly to take into account diverging phase gradients and vanishing order parameter amplitude at the cores of vortices or at nodes of phase slips, related strong electric fields and supercritical variations in concentration of normal electrons.

Related cases of current vortices in superconductors and electronic vortices in CDW are described usually within Ginzburg-Landau (GL) like models and their time-dependent (TDGL) generalizations for superconductors [35,36] and CDWs [37-40]. The essence of the GL and TDGL approach is that electrons are excluded ("fermions are integrated out") at the earlier stage so that the order parameter alone presents all observable quantities. We will show below that, while being consistent for superconductors, this approach fails for electronic crystals and this discrepancy is particularly pronounced for processes with a dynamical vorticity - phase slips or moving dislocations, leading to an unphysical production or elimination of condensed particles in their cores. We shall reconsider the theory working at all steps explicitly with interfering order parameter and normal electrons. Employing the chiral invariance and considering the related quantum anomaly allows to construct the treatable theory and perform the modeling without descending to the level of microscopic calculations. The price paid is that the theory is non-analytical with respect to the order parameter which originates some difficulties in numerical modeling. Nevertheless, with some simplifications we succeed to demonstrate the dynamical creation of vortices and phase slips processes.

The article is organized as follows: in Sec. 2 we discuss problems and contradictions appearing in treatment of the electronic crystals within the TDGL scheme; Sec. 3 is devoted to advantages and curiosities of chiral transformations and to the role of the chiral anomaly; in Sec. 4 we derive general equations for co-evolution of the order parameter, the normal carries and the electric field; in Sec. 5 we consider simplifications of limiting cases of the infinite normal conductivity and/or the local electro-neutrality; Sec. 6 is devoted to the numerical modeling, and conclusions are presented in Sec. 7.

## 2. Violation of charge conservation in the TDGL approach.

Phenomenological theories provide well established tools to describe classical and quantum coherent phenomena. They exploit basically the order parameter symmetry with some minimal simplifications. The most important symmetry classes are the complex order parameter describing superfluids, superconductors, incommensurate CDWs, the vector order parameter for magnetic skyrmions [41], and more complicated cases in He3 phases [42]. For the complex order parameter the most known advances of the symmetry approach include the Gross-Pitaevskii and the TDGL theories. Thus, for superconductors the theory of the vortex state was given by Abrikosov based on the GL approach while advances in theory of phase slips have been made based on the TDGL approach (see e.g. [35] and rfs. therein).

In the background of the GL, TDGL approaches, the order parameter $\eta = C exp(i\vartheta)$, composed with the amplitude $C$ and the phase $\vartheta$, is built from pairs of fermions at the microscopical level. But at the macroscopic level, the variational functional and resulting equations depend on $\eta$ alone with fermions being "integrated out". Perturbations of the fermionic system due to a finite temperature (*T*) or an out-of-equilibrium state affect the parameters like the equilibrium value of *C* or the collective ("superfluid") density $\rho_c$ but still they are not present explicitly. This procedure is sensitive with respect to the local charge conservation since the charge and the current densities are shared among collective and normal reservoirs. In the TDGL theory for superconductors, the charge conservation is automatically preserved because the equation of motion for the order parameter contains only gauge invariant combinations of the phase and the electromagnetic field. But in CDWs the order parameter phase characterizes breaking of the chiral (translational) rather than the gauge symmetry. The equations of motion for the CDW phase certify the local equilibrium of forces (elastic, Coulomb, frictional) with no relation to the charge conservation. Instead, the last is usually preserved by construction of the charge and current densities:

$$n_c = -\rho_c \partial_x \vartheta / \pi; \quad j_c = -\rho_c \partial_t \vartheta / \pi \qquad (1)$$

(to be compared with $j_{SC} \propto \rho_{SC} \partial_x \theta$ for the current dependence on the phase $\theta$ in superconductors). Here and below the currents refer to the number of particles and electronic charge e is incorporated into the corresponding potentials. The parameter $\rho_c(C)$ can be derived microscopically as a function of the equilibrium value of *C(T)* which itself takes into account the thermal population of normal carriers. It has a property that $\rho_c(1)=1$ and $\rho_c(C) \sim C^2 \to 0$ at $C \to 0$.

The expressions (1) ensure the charge conservation automatically indeed, but (!) only if the CDW amplitude is invariable: *C(t,x)=cnst*. Otherwise

$$\frac{dn_c}{dt} = \partial_t n_c + \partial_x j_c = \frac{1}{\pi}(\partial_x \rho_c \partial_t \vartheta - \partial_t \rho_c \partial_x \vartheta) \neq 0 \qquad (2)$$

We see from (2) that the charge conservation $dn_c/dt=0$ is violated in the current-carrying state $\partial_t \vartheta \neq 0$ if $\rho_c$ is not space homogeneous $\partial_x \rho_c \neq 0$ and in the strained (charged) state $\partial_x \vartheta \neq 0$ if $\rho_c$ varies in time $\partial_t \rho_c \neq 0$: the unphysical production/elimination of condensed particles takes place, thus violating the gauge invariance. The effect is particularly disturbing in description of motion or nucleation of vortices or of phase-slip processes when space and time derivatives are high near the vortex core or the amplitude node. It looks that for a general nonlinear spacio-temporal regime there is no explicit way to define the CDW collective charge and current via the order parameter alone. In Sec. 4 we shall see that, following the evolution and deformations of the phase, the CDW current $j_c$ and the density $n_c$ are given by the nominal (zero temperature) contribution from the total number of electrons independently of the thermal or non-equilibrium occupation of excited states. The total charges and currents are additive

$$n_{tot} = n_c + n_n = \frac{1}{\pi} \partial_x \vartheta + n_n \quad j = j_c + j_n = -\frac{1}{\pi} \partial_t \vartheta + j_n \qquad (3)$$

where the collective ($n_c$, $j_c$) and normal ($n_n$, $j_n$) components satisfy the charge conservation law separately: the first one by construction, the second one by definition. Nevertheless, the reduced quantities $n_{tot} = \rho_c \partial_x \vartheta / \pi$ and $j_{tot} = -\rho_c \partial_t \vartheta / \pi$ $\pi$ do appear under certain conditions with $\rho_c = 1 - \rho_n$ where the compensation $\rho_n$ comes from the drag $\partial_x \vartheta/\pi$ and effect of the variable phase upon normal electrons.

To resolve this problem, it is ultimately necessary to take into account the normal carriers explicitly, without integrating them out. At first sight, that would require for descending to the fully microscopic theory with its notorious complications even in linear or gapless regimes [32,37]. From experience of superconductors, the game among the collective and the normal subsystems gives rise to problems undermining the very existence of the TDGL approach (see discussion and references in [35]. In the next section we shall nevertheless demonstrate the way how to keep the phenomenology based on the knowledge of chiral invariance and transformation, impotently taking into account the chiral anomaly.

Finally notice that the paradox is present in theory of an arbitrary electronic crystal beyond the quasi one-dimensionality and particular microscopics inherent to CDW. E.g. for a Wigner crystal the charge and the current densities are expressed via the field of displacements **u(r,t)** as

$$n_c = -\nu \nabla \mathbf{u}; \quad \mathbf{j}_c = \nu \partial_t \mathbf{u}$$

where we have introduced the filling factor ν(**r**,t) of the crystal unit cell by the electron. Then

$$\frac{dn_c}{dt} = \partial_t n_c + \nabla \mathbf{j}_c = \nabla \nu \partial_t \mathbf{u} - \partial_t \nu \nabla \mathbf{u} \neq 0$$

hence the conservation law is violated for variable ν.

### 3. Normal carriers: paradoxes resolved by the chiral anomaly.

We proceed within the common picture of a weak CDW $\Delta \ll E_F$ which allows to decompose 1D electrons in right and left moving components near Fermi points $\pm P_F$ and to linearize their spectra keeping the CDW profile with wave numbers only close to $2P_F$:

$$\Psi = \psi_+ \exp(iP_F x) + \psi_- \exp(-iP_F x) \text{ , and CDW modulation is } \propto \Delta \cos(2P_F x + \vartheta).$$

That gives rise to the pseudo-spinor of right and left moving components $(\psi_+, \psi_-)$ and to the order parameter as $\eta = Ce^{i\vartheta}$, where $C = \Delta/\Delta_0$ is the ratio of the CDW local gap value to its $T=0$ equilibrium value $\Delta_0$.

Following the principles of the Peierls-Froehlich model [2], we take into account interactions of electrons with the CDW and with the electromagnetic field. Implicitly, we shall allow also for scattering of electrons by CDW excitations (collisions with other electrons, phase and amplitude modes). We do not take into account the scattering in the frame of the host lattice which brings dependencies on the CDW phase: no impurities, Umklapp processes, or commensurability effects (for a homogeneous state or within the linear response the theory of these effects has been developed with tiny details, particularly in [31-34]).

The Lagrangian $H - i\hbar \partial_t$ for electrons has a form:

$$\mathcal{L} = \begin{pmatrix} -i\hbar\partial_t - i\hbar v_F \partial_x + \Phi - v_F A_x & \Delta e^{i\vartheta} \\ \Delta e^{-i\vartheta} & -i\hbar\partial_t + i\hbar v_F \partial_x + \Phi + v_F A_x \end{pmatrix} \quad (4)$$

where $\Phi$ and $A_x$ are the scalar and the vector potentials (incorporating factors $e$ and $e/c$ correspondingly). We perform the chiral transformation which brings the wave function to the local frame of a distorted CDW phase:

$$\psi_\pm \rightarrow \psi_\pm \exp(\pm i\vartheta/2) \quad (5)$$

$$\mathcal{L} \Rightarrow \begin{pmatrix} -i\partial_t - i\hbar v_F \partial_x + \Phi - v_F A_x + \frac{\hbar v_F}{2}\partial_x \vartheta + \frac{\hbar}{2}\partial_t \vartheta & \Delta \\ \Delta & -i\partial_t + i\hbar v_F \frac{\partial}{\partial x} + \Phi + v_F A_x + \frac{\hbar v_F}{2}\partial_x \vartheta - \frac{\hbar}{2}\partial_t \vartheta \end{pmatrix} \quad (6)$$

Now the gap term loses the dependence on the phase, but in expense of additional parts in potentials $\Phi$ and $A_x$:

$$\Phi \rightarrow V = \Phi + \frac{\hbar v_F}{2}\partial_x\vartheta \qquad A_x \rightarrow A = A_x + \frac{\hbar}{2v_F}\partial_t\vartheta \qquad (7)$$

The additions to $\Phi$ and $A_x$ from $x,t$ derivatives of the phase are naturally interpreted as the Fermi energy shift $\hbar v_F \delta P_F$ following the Fermi momentum shift $\delta P_F = \partial_x\vartheta/2$. under the CDW phase deformations. Chiral invariant combinations (7) provide effective scalar and vector potentials $V$ and $A$ yielding the total longitudinal force $E_x^*$ experienced by electrons under the CDW phase deformation and the applied electric field $E_x$:

$$E_x \rightarrow E_x^* = -\partial_x V + \partial_t A = E_x - \frac{\hbar v_F}{2}\left(\frac{\partial^2\vartheta}{\partial x^2} - \frac{1}{v_F^2}\frac{\partial^2\vartheta}{\partial t^2}\right) \qquad (8)$$

At first sight, we have arrived just at the transparent picture of a 1D semiconductor with the gap $2\Delta$ under the effective electric field (8). Now it looks straightforward to exclude the fermions to arrive at an effective action $S\{\Delta, \vartheta, \Phi, A\} = \int\!\!\!\int W dx dt$. And here we come to puzzling contradictions described below:

i. Let us concentrate only on effects of $V$ and keep $\Delta$ =cnst for a while. In view of Eq. (7), the free energy $W$ should contain the potential and the phase only in the invariant combination $V$. Since we can always choose the phase $\vartheta$ such that $V = \Phi + \frac{\hbar v_F}{2}\partial_x\vartheta = 0$, then the applied potential disappears from the action, hence no density and no polarization are perturbed with respect to $\Phi$ which contradicts to basic properties of the CDW as both the semiconductor with respect to electrons and the metal in the collective behavior.

ii. Consider now effects of $\partial_x\vartheta$ and $\Phi$ separately keeping in mind that according to (7) they should be interchangeable. From the general experience of systems with the symmetry breaking with respect to the complex order parameter, we expect $W$ to contain the term $\propto \rho_c(\partial_x\vartheta)^2$ where $\rho_c$ is a collective density responsible for the phase rigidity. But treating the Lagrangian (6) perturbatively with respect to the effective electric field (8) we evidently should get

$$\delta W_\epsilon = -\frac{s_\perp \varepsilon(k,\omega)}{8\pi e^2}(E_x^*)^2 = -\frac{s_\perp \varepsilon(k,\omega)}{8\pi e^2}\left(-\frac{\hbar v_F}{2}\frac{\partial^2\vartheta}{\partial x^2} + E_x\right)^2 \qquad (9)$$

where $s_\perp$ is the perpendicular area par chain and $\varepsilon$ is the dielectric susceptibility as a function of the wave number $k$ and the frequency $\omega$ (the last will be neglected here for shortness). At small $k$:

$$\epsilon = \epsilon_\Delta + \frac{1}{(lk)^2}, \frac{1}{l^2} = \frac{\rho_n}{r_0^2}, \rho_n = \frac{dn}{d\zeta}N_F^{-1} \qquad (10)$$

Here $r_0$ is the Thomas-Fermi radius of the parent metal, $n$ is the concentration of free electrons depending on their chemical potential $\zeta$ and $\rho_n$ is the normalized DOS which also will happen to be the normal density. The dielectric constant $\epsilon_\Delta \propto (r_0\zeta_0)^{-2}$ collects the polarization of electrons gapped by the CDW while $\rho_n$ comes from conducting carriers thermally excited or injected [7] above the gap providing the finite screening length $l\sqrt{\epsilon_\Delta}$. Both contributions bring drastic contradictions:

i. The term with $\epsilon_\Delta$ in Eq. (10) yields to (9) the forth order gradients of the phase $(\partial_{xx}^2\vartheta)^2$ instead of the second order $\propto (\partial_x\vartheta)^2$ expected for the elastic energy of CDW deformations.

ii. The singularity in $k$ in the term with $\rho_n$ seems at first to serve fortunately by cancelling excess gradients leading to the contribution $\propto (\partial_x\vartheta)^2$

$$\delta W_{\rho n} = -\rho_n \frac{\hbar v_F}{4\pi}\left(\partial_x\vartheta + \frac{2}{\hbar v_F}\Phi\right)^2 \qquad (11)$$

but with two confusions with respect to the expected

$$W_{\rho c} = \rho_c \frac{\hbar v_F}{4\pi}(\partial_x\vartheta)^2 \qquad (12)$$

- even the signs in Eqs. (11) and (12) are opposite, and temperature dependencies are conflicted among coefficients $\rho_n$ (expected to rise from zero at $T=0$ up to 1 at $T_c$) and $\rho_c$ (expected to fall to zero at $T_c$ starting from 1 at $T=0$).

These contradictions can be traced back to the notion of the chiral anomaly and they can be cured by properly taking this anomaly into account. The chiral anomaly is ubiquitous to the premature linearization of electronic spectra. First, in the course of the linearization the control is lost of the position of the bottom of the electronic band, hence of the difference between action of the external potential and of the Fermi energy shifting. Moreover, the whole energy cost of the chiral transformation $\delta W_{CA}$ is lost. This energy can be captured from the non-linearized spectrum of the normal metal if we consider $\delta n_{CT} = \partial_x \vartheta / \pi$ as a redistribution of the total particle density accompanying the chiral transformation

$$\delta W_{CA} = \frac{\hbar v_F}{4\pi}(\partial_x \vartheta)^2 + \frac{\Phi}{\pi}\partial_x \vartheta \qquad (13)$$

where the first term is the density perturbation cost $(\delta n_{CT})^2/(2N_F)$ ($N_F$ is the DOS of the parent metal) and the second term is its potential energy $\delta n_{CT}\Phi$. Beyond these physical arguments [29,30], the derivation of the chiral anomaly in the spirit of the field-theory procedure of regularization of fermionic determinants was demonstrated for CDWs [43] and for SDWs [44].

Bringing together the nonperturbative contribution (13) and the perturbative one (11), we get

$$W_{tot} = \delta W_{\rho n} + \delta W_{CT} = \rho_c \left(\frac{\hbar v_F}{4\pi}(\partial_x \vartheta)^2 + \frac{\Phi}{\pi}\partial_x \vartheta\right) - \rho_n \frac{(\Phi)^2}{\pi \hbar v_F}, \quad \rho_c = 1 - \rho_n \qquad (14)$$

which correctly manifests the expected properties (12) and (1) yielding also the important relation $\rho_c + \rho_n = 1$. Even more general relation among the collective and normal responses follows: $\rho_n = \epsilon(k,\omega) r_0^2 (k^2 - \omega^2)$, whatever is the dielectric function of electrons in the CDW semiconductor. The total charge density becomes

$$n_{tot} = \frac{\delta W_{tot}}{\delta \Phi} = \frac{1}{\pi}\rho_c \partial_x \vartheta - \rho_n \Phi N_F = \frac{1}{\pi}\partial_x \vartheta - \rho_n V N_F$$

Here the first form of $n_{tot}$ interpret the CDW charge as $\rho_c$ leaving the electronic density to react only to $\Phi$ which approach is common and convenient while ambiguous: with a variable $\rho_c$ this form leads to violation of the charge conservation as it was discussed in the previous section. The second form of $n_{tot}$ lets us to understand that the collective charge density is always the nominal one embracing all electrons, not reduced by the factor $\rho_c$; meanwhile the charge density of normal electrons is perturbed by their actual combined potential $V$ rather than by its part $\Phi$. In this picture, the charge and the current are given by independently conserved counterparts as show in Eq. (3).

The general picture is that the proper energy, charge (and hence the current) of the total system are given by the total number of electrons as for $T=0$. But the reaction of conducting electrons provides a compensating contribution $\sim \rho_n$ which acts to erase the bare quantities reducing their magnitude leading to the factor $\rho_c$. This compensation can be seen explicitly only for this simple case of the linear response theory with the constant amplitude. But the provided basis will be exploited below to build the general nonlinear scheme for all variables.

Technically (see [18]), the anomaly appears from uncertainty in calculation the Fermionic determinants which must be regularized at some lattice which procedure is usually treated in a relativistically invariant way in traditions of the field theory. It is instructive to see the origin of the anomaly in a transparent way related to rules of many-electronic systems, namely to the Tomas-Fermi procedure. Let the electrons in the parent metal occupy a parabolic spectrum $p^2/2m$ up to some Fermi energy $E_F$. Recall the manual book procedure: in an arbitrary smooth potential $V(x)$ the electronic wave functions and their density distributions for an eigenenergy $E$ are legitimately given by the WKB expressions

$$\Psi_E \propto (E - V(x))^{-1/4} \exp\left(i \int dx' (E - V(x))^{1/2}\right), \quad \rho_E = |\Psi(x)|^2 = (E - V(x))^{-1/2}$$

Summing up $\rho_E$, the total density can be obtained and expanded in $V/E_F$ as $\rho(x) \approx <\rho> + V(x)N_F$ which yields the most important characterization of a metal: the linear response of the density to the applied potential. The DOS factor $N_F$ reminds that the important states were those near the Fermi points where the spectrum linearization looks to be legitimate. But if this linearization is done prematurely, then the Schrödinger eq. becomes the first order one, the wave function $\Psi_E$ looses the prefactor, $\rho(x)$ becomes just constant, and the reaction to the local potential disappears. The role of the chiral anomaly action (W-CA) is just to restore this missing contribution.

### 4. General equations for nonlinear regimes.

The CDW conductor maintains in general two different types of normal carriers, extrinsic and intrinsic ones. The extrinsic carriers with a concentration (henceforth per unit length of one chain) $n_{ext}$ can be originated from left open parts of the Fermi surface or from the bands which do not participate in formation of the CDW; they do not interact with CDW directly, but only via common electric potential $V_{ext}=\Phi$. Their spectrum is unaffected by the CDW gap and no forces are coming from the CDW phase deformations (a "Dark Matter" analog in condensed matter).

Intrinsic carriers belong to the bands forming the CDW and they can be activated across the CDW gap; those we kept in mind in discussions of previous sections. In addition to the Coulomb interactions, intrinsic carriers are affected also by the CDW deformations according to Eqs. (7,8), which are coming from local shifts of their Fermi energy and momentum taken into account by the chiral transformation of Eqs. (5,6). The total potential energy $V$ for intrinsic electrons has the form of Eq. (7).

We shall use basic parameters of the parent metal and of the equilibrium CDW: $\varepsilon_{host}$ as the dielectric susceptibility of the host lattice, Fermi velocity in the parent metal $v_F$, DOS per chain $N_F$, Thomas-Fermi screening length $r_0$, the CDW gap $\Delta_0$ and the coherence length $\xi_0$:

$$N_F = \frac{2}{\pi \hbar v_F} \, , \, \frac{1}{r_0^2} = \frac{4\pi e^2 N_F^2}{s\varepsilon_{host}} = \frac{8e^2}{s\varepsilon_{host}\hbar v_F} \, , \, \xi_0 = \frac{1}{N_F \Delta_0}$$

In or near the equilibrium, the thermodynamics is governed by the free energy $F(C,n_e,n_h)$ as a function of the normalized gap value $C=\Delta/\Delta_0$ and of the concentration (per unit length of the chain) of intrinsic normal carriers among which we separate electrons $n_e$ and holes $n_h$ with the notation $n=n_e-n_h$. In presence of extrinsic carriers, we should add also their free energy $F_{ext}(n_{ext})$. The equilibrium value $C_{eq}$ is connected with $n_e,n_h$ via the minimum, $\partial F(C,n_e,n_h)/\partial C=0$, in such a way that $C_{eq}(n_e,n_h)$ vanishes when $n_{e,h}$, or better say their chemical potential $\zeta$ surpass critical values, hence the metallic phase with $C=0$ is restored.

The total energy functional $\int W dx$ depends on densities of intrinsic $n$ and extrinsic $n_{ex}$ carriers, the electric potential $\Phi$, the phase and the amplitude of the order parameter. The energy density (per unit length of one chain) is

$$W = \frac{\hbar v_F}{4\pi}\left[\kappa_x(\partial_x C)^2 + \kappa_y(\partial_y C)^2 + \kappa_z(\partial_z C)^2 + (\partial_x \vartheta)^2 + \kappa_y C^2 (\partial_y \vartheta)^2 + \kappa_z C^2 (\partial_z \vartheta)^2\right] +$$

$$\frac{1}{\pi}\Phi \partial_x \vartheta + \left(\Phi + \frac{\hbar v_F}{2}\partial_x \vartheta\right)n + \Phi n_{ext} - \frac{\varepsilon_{host} s}{8\pi}(\nabla \Phi)^2 - \frac{\varepsilon_\Delta s}{8\pi}(\nabla V)^2 + F(C,n_e,n_h) + F_{ext}(n_{ex}) \quad (15)$$

Here parameters $\kappa_y,\kappa_z$ describe the interchain coupling of CDWs, and we shall put the on-chain rigidity of the amplitude $\kappa_x \sim 1$ to be $\kappa_x=1$ in correlation with the phase rigidity term $(\partial_x \vartheta)^2$ - in the spirit of the GL approach. The terms with $\sim (\partial_x \vartheta)^2$ and $\sim \frac{1}{\pi}\Phi \partial_x \vartheta$ are originated by the chiral anomaly of Eq. (13) coming from the background deformations of the CDW phase, the term $\sim n$ comes from the energy of intrinsic electrons in the combined potential $V = \Phi + \frac{\hbar v_F}{2}\partial_x \vartheta$. The derivative of $W$ over the CDW strain $\partial_x \vartheta/\pi$ gives the CDW longitudinal stress

$$U = \frac{1}{\pi N_F}\partial_x \vartheta + \Phi + \frac{n+n_{ex}}{N_F} = V + \frac{n+n_{ex}}{N_F}$$

which can be also interpreted as the chemical potential of condensed electrons $\mu_c=U$.

Assuming the dissipative regime for both $\vartheta$ and $C$, functional derivatives of (15) yield equations for their time evolution:

$$\partial_x(\partial_x\vartheta + \pi N_F \Phi + \pi n) + \kappa_y \partial_y(C^2 \partial_y \vartheta) + \kappa_z \partial_z(C^2 \partial_z \vartheta) = \gamma_\vartheta \partial_t \vartheta \qquad (16)$$

$$\kappa_x \partial_x^2 C + \kappa_y \partial_y^2 C + \kappa_z \partial_z^2 C + \kappa_y C(\partial_y \vartheta)^2 + \kappa_y C(\partial_z \vartheta)^2 - \partial F/\partial C = \gamma_C \partial_t C \qquad (17)$$

Here $\gamma_\vartheta = \gamma C^2$, $\gamma$=cnst, $\gamma_C$=cnst are the CDW phase and amplitude damping coefficients correspondingly. $\gamma_\vartheta$ is related to the sliding CDW conductivity $\sigma_{CDW}$ as $\gamma_\vartheta \sigma_{CDW} = N_F e^2/s = 1/(4\pi r_0^2)$. Variation $\delta W/\delta \Phi = 0$ yields the Poisson equation for $\Phi$ (neglecting higher derivatives of the phase):

$$N_F r_0^2 \nabla^2 \Phi + \frac{1}{\pi}\partial_x \vartheta + n + n_{ex} = 0 \qquad (18)$$

Contrary to conventional GL-type equations, now Eqs. (16-18) are nonanalytic in the order parameter $\eta = C \exp(i\vartheta)$ or in other words singular in its amplitude. The terms containing $\partial_x \vartheta$, $\partial_{xx}\vartheta$ and $(\partial_x \vartheta)^2$ are not multiplied by $C^2$, so the attempt to present them as derivatives of $\Psi$ will bring $\sim 1/C, 1/C^2$ singularities. Nevertheless, there are hidden cancelations allowing to compensate for singularities, even if implicitly, which is vitally important for allowance of space- and spacio-temporary vortices. In the following sections, with the help of additional approximations, these compensations will be better exposed or proved at least.

Eqs. (16-18) need to be complemented by an equation for the normal density $n$. Even if the chiral transformation has reduced the task to electrons of a semiconductor with a gap $2\Delta$ subjected to the effective potential $V$, the detailed description of electrons' kinetics is still a complicated problem taking into account different collisions and scattering mechanism and particularities of (quasi) one-dimensionality. In the context of the CDW, the most systematic studies have been performed in series of articles [32 and rfs. therein, 33,34]. Results of this theory show various regimes at low $T$ and at $T\sim T_c$ including the gapless regime [37-39] which is an analog to superconductors with paramagnetic impurities. The last case of dirty materials (never exploited or even reached in experiments) is the only one where our approach may not be applicable even qualitatively since this regime is dominated by the backscattering processes which we do not take into account. (Their contributions $\propto \psi_+^\dagger \psi_- \exp(i\vartheta + iQr_i)$ to the Hamiltonian depend on the chiral phase explicitly.) Because of the positional randomness of phase shifts $Qr_i$, the chiral invariance is restored in average, but at expense of disturbing the distribution function of electrons by the time dependence of $\vartheta$. That was shown [45] to bring about the countercurrents of normal electrons $\delta j_{cc} \sim \partial_t \vartheta$ which effect can be taken into account supplementary in our approach. With these reservations, the kinetics of normal carriers will be taken in the quasi-equilibrium diffusive approximation:

$$\nabla(\hat\sigma \nabla \mu) = \frac{e^2}{s}\partial_t n \; ; \; \mu = \zeta + \Phi + \frac{1}{\pi N_F}\partial_x \vartheta \; ; \; \hat\sigma = e\hat b(n_e + n_h) \qquad (19)$$

$$\nabla(\hat\sigma_{ex} \nabla \mu_{ex}) = \frac{e^2}{s}\partial_t n_{ex} \; ; \; \mu_{ex} = \zeta_{ex} + \Phi \; ; \qquad \hat\sigma_{ex} = e\hat b_{ex} n_{ex}$$

Here $b$ and $b_{ex}$, $\mu$ and $\mu_{ex}$, $\zeta = \partial F/\partial n$ and $\zeta_{ex} = \partial F_{ex}/\partial n_{ex}$ are local mobilities, chemical and electrochemical potentials of intrinsic and extrinsic carriers, $\hat\sigma = (\sigma_x, \sigma_y, \sigma_z)$ is the anisotropic conductivity tensor. The equilibration between the normal particles is expected to be very fast, hence their electrochemical potentials can be treated as identical: $\mu_{ex} = \mu$, then $\zeta_{ex} = \zeta + \frac{1}{\pi N_F}\partial_x \vartheta$, and finally

$$\nabla\left(\hat\sigma_t \nabla(\zeta + \Phi + \frac{1}{\pi N_F}\partial_x \vartheta)\right) = \frac{e^2}{s}\partial_t(n + n_{ex}); \quad \hat\sigma_t = \hat\sigma + \hat\sigma_{ex} \qquad (20)$$

Instead of $n$, we can use the local chemical potential $\zeta$ related to $n$ by the equation of state $n(\zeta,C)$. This dependence also defines dimensionless normal and collective densities $\rho_n$ and $\rho_c$:

$$\zeta = \frac{dF}{dn}, \qquad \rho_n = \frac{1}{N_F}\left(\frac{dn}{d\zeta}\right)_C, \quad \rho_c = 1 - \rho_n$$

In the metallic phase $\rho_n=1$, then $\rho_c=0$ growing inwards the CDW phase as $\rho_c \sim C^2$ reaching $\rho_c=1$ in equilibrium at $T=0$. Notice that the difference of electrochemical potentials of normal ($\mu_n=\mu$) and condensed ($\mu_c=U$) electrons

$$\delta\mu = \mu_n - \mu_c = \zeta - \frac{n+n_{ex}}{N_F} \tag{21}$$

measures the degree of non-equilibrium with respect to the conversion among the two reservoirs [6-8] which might be very slow requiring for phase-slip processes.

The extrinsic carriers do not interact with CDW directly, rather only via the Coulomb potential, their contribution can be added posteriori, so since now on we shall omit $n_{ex}$ in the equations.

## 5. Simplifying equations in limiting cases.

### 5.1. Infinite conductivity of intrinsic carriers.

The limit of the infinite conductivity of intrinsic carriers makes sense at least for 1D regimes at sufficiently high $T$ when only the possibly high $\sigma_x$ is involved. Now the electrochemical potential $\mu$ must be constant which we can put as zero:

$$\mu = \zeta + \Phi + \frac{1}{\pi N_F}\partial_x\vartheta = 0 \tag{22}$$

Excluding $\Phi$ from Eqs. (16,18) with the help of Eq. (22) we arrive at

$$\frac{\partial}{\partial x}(n/N_F - \zeta) + \kappa_y\partial_y(C^2\partial_y\vartheta) + \kappa_z\partial_z(C^2\partial_z\vartheta) = \gamma_\vartheta\partial_t\vartheta \tag{23}$$

$$r_0^2\nabla^2(N_F\zeta + \partial_x\vartheta/\pi) = (\partial_x\vartheta/\pi + n + n_{ex}) \tag{24}$$

The behavior of different terms in the Eq. (23) at small $C(t,x)$ is consistent since at $C\to 0$ $\zeta$-$n/N_F\to 0$ by construction and $\gamma_\vartheta \to 0$ by definition. Curiously, the Eq. (23) does not show explicitly the commonly assumed longitudinal phase rigidity $\propto \partial_x^2\vartheta$; it is hidden in the first term $\frac{\partial}{\partial x}(\pi n - \zeta)$ implicitly, via Eq. (24). This term provides also the driving force for the CDW current which can be written as $-\rho_c\partial_x\zeta$ instead of the conventionally supposed $\rho_c E$: gradient of the normal carriers concentration rather than the electric field. The relation (22) helps also to understand the decomposition of the total charge density $n_{tot}$ as it is given in the RHS of Eq. (24). We can write

$$\partial_x n_{tot} = -\varrho_n N_F \partial_x\Phi + \frac{\varrho_c}{\pi}\partial_x^2\vartheta$$

In the RHS of this expression the first term corresponds, for $\rho_n = cnst$, to the conventional reaction (the screening of the electric field with a local screening length $l_{scr}^2=r_0^2/\rho_n$) of normal carriers to the applied electric field as if the CDW phase is not deformed. The second term offers an interpretation of an effective collective charge density given by its derivative: $\partial_x n_c^{**} = \rho\partial_x^2\vartheta/\pi$ which coincides with the commonly expected $n_c^* = \varrho_c\partial_x\vartheta/\pi$ only if $\rho_c = const$.

### 5.2. Local electroneutrality.

Formally the approximation of the local electroneutrality corresponds to the limit $r_0=0$ in the Eq. (18) which is fully justified for a typical dense crystal of chains where $r_0 \sim$ interatomic distances. Now the Eq. (18) is reduced to the relation

$$\partial_x\vartheta + \pi n = 0, \text{ then } U = \Phi, \mu = \Phi + \zeta - n/N_F \tag{25}$$

and the Eq. (16) acquires the form

$$\partial_x\Phi + \frac{1}{\pi N_F}\left(\kappa_y\partial_y(C^2\partial_y\vartheta) + \kappa_z\partial_z(C^2\partial_z\vartheta)\right) = \gamma_\vartheta\partial_t\vartheta \tag{26}$$

Curiously again, the Eq. (26) does not show explicitly the commonly assumed phase rigidity $\partial_{xx}^2 \vartheta$ along the chains: the electric field drives the collective current but does not deform the CDW, the last function is left for *n*, according to Eq. (25).

The diffusion Eq. (20) for normal carriers becomes

$$\nabla(\sigma \nabla(\Phi + \zeta - nN_F)) = \frac{e^2}{s} \partial_t n \qquad (27)$$

The driving field $-\nabla(\Phi + \zeta - n/N_F)$ for the normal current becomes $-\nabla\Phi$-$(\rho_c/\rho_n) \nabla n$ (if *C=cnst*) i.e. the effective diffusion coefficient is enhanced as $\rho_c/\rho_n$ which result could hardly be expected intuitively. Eqs. (26) and (27) with the relation (25) constitute the complete system of equations for $\vartheta$, $\Phi$, *n*. In the 1D regime we can use Eq. (25) to integrate Eq. (27) over *x* and then exclude $E_x$ with the help of Eq. (26) to arrive at the Eq. for the phase alone:

$$\frac{1}{\pi}\left(\frac{1}{\sigma_{CDW}} + \frac{1}{\sigma_n}\right)\partial_t \vartheta + s\partial_x(\frac{1}{\pi}\partial_x \vartheta + \zeta) = \frac{-1}{\sigma_n} J(t) \qquad (28)$$

Here explicit functions of *n* and *C*: $\zeta(n,C)$, $\sigma_n = b(C)n_{tot}(\zeta,C)$, $\sigma_{CDW} \propto 1/C^2$ are exploited as functions of *C*, $\partial_x \vartheta$ via Eq. (25). The RHS of Eq. (28) gives the driving force as the total current *J(t)* multiplied by the normal resistance alone. The second term in the LHS describes the phase rigidity; at *C=cnst* it can be written as $\frac{\rho_c}{\rho_n}\partial_x^2 \vartheta$ where the ratio $\rho_c/\rho_n$ controls vanishing of the rigidity at *C→0* approaching the metallic state and the Coulomb hardening [32,46,47] of the charged phase deformations which dramatically increases with freezing out of screening by normal carriers when $\rho_n \to 0$.

### 5.3. Local electroneutrality together with the infinite conductivity.

Now we join both limits: the local electroneutrality, $r_0=0$, and the infinite conductivity of intrinsic carriers, µ=0. These conditions allow to express $\Phi, \partial_x \vartheta, U$ via $\zeta$ or equivalently via *n* alone:

$$U = \Phi = n/N_F - \zeta, \quad \partial_x \vartheta = -\pi n(\zeta, C) \qquad (29)$$

Then we arrive at two equivalent forms of equations for the phase:

$$-N_F \partial_x(\zeta - n) + (\kappa_y \partial_y C^2 \partial_y + \kappa_z \partial_z C^2 \partial_z - \gamma_\vartheta \partial_t)\vartheta/\pi = 0 \qquad (30)$$

$$\frac{\rho_c}{\rho_n}\partial_x^2 \vartheta + (\kappa_y \partial_y(C^2 \partial_y \vartheta) + \kappa_z \partial_z(C^2 \partial_z \vartheta) - \gamma_\vartheta \partial_t)\vartheta = \pi N_F \frac{\partial \zeta}{\partial C}\partial_x C \qquad (31)$$

The form (31) shows explicitly the physical phase rigidity accumulating all effects of normal carriers and Coulomb interactions. Remarkably, in this form the Eq. for $\vartheta$ does not show any driving force, apart from the less important term in the RHS coming from the gradient of the amplitude *C*. The drive will come only from the boundary conditions for the electric potential transferred to the phase via relations (29). For the 1D regime, the Eq. (30) agrees with the Eq. (28) in the limit $\sigma_n \to \infty$.

In the 1D regime we can exclude the phase by differentiating the Eq. (30) over *x* to arrive at the closed eq. for only one (apart *C* which is governed by Eq. (17) for all cases) variable $\zeta$ or *n*:

$$\gamma^{-1}\partial_x(C^{-2}\partial_x(N_F \zeta - n)) = \partial_t n \qquad (32)$$

If necessary, the phase can be restored by integration of $\partial_x \vartheta = \pi n$.

### 6. Numerical modeling

Equations derived above for the general case or for specific limits can be used for numerical modeling. The equations written transparently in terms of the phase and the amplitude must be implemented in terms of periodic functions of the phase - the complex order parameter or its components: $\eta = C\exp(i\theta)$. Otherwise, in terms of the phase obtained as a single-valued function, all topologically nontrivial solutions (dislocations as space vortices and phase-slips as space-time vortices) will be missed. In the above written eqs, the phase derivatives are treated as local ones: $\partial_t \vartheta \to \omega = Im(\eta * \partial_t \eta)$, $\partial_x \vartheta \to q = Im(\eta * \partial_x \eta)$. The phase can be restored as $\vartheta \to \vartheta^t$ or $\vartheta \to \vartheta^x$ by integrating $\omega$ over *t* or *q* over *x* correspondingly. The two definitions of the phase $\vartheta^t$ and $\vartheta^x$ are not identical in presence of the space-time vorticity: $(\partial_x\omega - \partial_t q)/2\pi = R$ is the density of the vortex production rate, $(\partial_x\vartheta^t - q)/2\pi = N$ is the vorticity density or the vortex contribution to the CDW strain.

We are dealing with partial differential equations which numerical solution requires for initial and boundary conditions. The initial ones are naturally written as $n(x,0) = 0$, $\Phi(x,0) = 0$, $\vartheta(x,0) = 0$, $C(x,0) = C_{eq}$. Boundary conditions for $C$ are either $C = C_{eq}$ or $\nabla_n C = 0$ where $\mathbf{n}$ is the normal to the boundary. Eqs. For $\vartheta$ and $n$ have a form $\partial_t f - \nabla \mathbf{j}_f = 0$; then the boundary conditions are given for normal components of the "currents" $n\mathbf{j}_f$. The practically most important limit of the electro-neutrality of Eqs. (25,26) is singular with respect to general boundary conditions, particularly concerning the integrated 1D form of Eq. (28): the former boundary conditions for $j_n$ now contain the second derivative of the variable $\vartheta$ which is not allowed. We shall treat Eq. (28) over a length interval *(0<x<L)* with boundary conditions fixing the phase thus imitating the pinning at the sample boundaries: $\vartheta(0,t) = \vartheta(L,t) = 0$. Then the CDW current vanishes, $\partial_t\vartheta(x,0) = \partial_t\vartheta(x,L) = 0$, and the function *J(t)* becomes the monitored source/drain normal current.

There are some technical challenges in numerical implementations which one commonly does not meet in conventional GL approaches, see e.g. [23,48]. First is the control of compensations at $C \to 0$ in expressions for total charges, currents, and the condensate energy bringing to action the hidden function of the condensate density $\rho_c$. Second is the entanglement in dependences of thermodynamic functions and their derivatives on $C$ and on $n$ or $\zeta$: approaching of $n$ or $\zeta$ to critical values should eliminate the energy minimum over $C$ at $C \neq 0$ opening the metallic state, e.g. in the vortex core. Third, the equations are simpler and better treatable in variables $n$ rather than $\zeta$ while commonly we possess the Gibbs energy $\Omega$, $n$, and $\rho_n$ as functions of $\zeta$, $C$ rather than $F$, $\zeta$, and $\rho_n$ as functions of $n$, $C$ and the inversion of dependencies on $n$ and $\zeta$ can be done also only numerically.

The best, and may be the only analytically transparent, advancing is possible with the simplest Landau type expression for *F*:

$$F(n,C) = n^2/(2N_F) + (-\tau + (a\xi_0 n)^2)(C\Delta_0)^2 N_F/2 + bC^4\Delta_0^2 N_F/4, \quad a,b \sim 1 \quad (33)$$

$$\partial F/\partial C = C\Delta_0^2 N_F(-\tau + (a\xi_0 n)^2 + bC^2)$$

$$\zeta = (n/N_F)(1 + (aC)^2); \quad \rho_n = 1/(1 + (aC)^2); \quad \rho_c = (aC)^2/(1 + (aC)^2)$$

Formally it is valid at *T* close to the mean-field transition point, i.e. at $\tau = 1 - T/T_c \ll 1$ but we consider it as a commonly used parametrization even beyond small $\Delta$ allowing to consider normal carriers in the frame of the effective semiconductor as described in Secs. 3,4. In Eq. (33) the first term of the zero order in *C* is the contribution of the normal metal, while two other terms give the Landau type expansion in the order parameter with the proximity $\tau \propto (1 - T/T_c)$ to the transition temperature being shifted by presence of the normal carriers which critical concentration is $n_{cr} = 1/(a\xi)$ with the correlation length defined as $\xi = \xi_0/\sqrt{\tau}$.

It is expected that the normal conductivity is proportional to the total number of electrons and holes ($\sigma_n \propto n_{tot} = n_e + n_h$) while the eqs. operate with $n = n_e - n_h$. We can reasonably guess the relation $n_{tot} = \sqrt{n_{eq}^2 + n^2}$ which is rigorous for the activational behavior $n_{e,h} \propto \exp[(-\Delta \pm \zeta)/T]$ and satisfies the correct limit $n_{tot} \to |n|$ at large deviations from the equilibrium value $n_{eq}$.

An example of calculations based on Eqs. Eq. (17,28,33) is presented below for a sample of a length *L=100ξ* with the source and the drain of the current around points *x=20ξ* and *x=80ξ* distributed within narrow intervals *δx=0.1ξ*. The phase slips appear with no threshold, unlike in other published modelings; just the initiation time for the first phase slip increases with the diminishing current. We observe a strictly periodic, with intervals *~1/J(t)*, sequence of space-time vortices which are clearly identified by the zero node of the amplitude (Fig. 1) and by $2\pi$ steps of the phase (Fig. 2a) – both are centered at the current terminals.

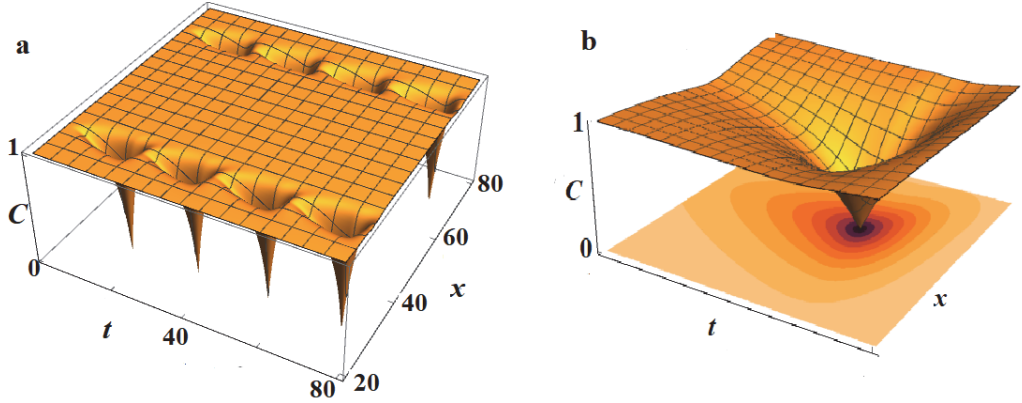

Fig. 1. 3D plots for dependencies of the amplitude $C(x,t)$ of the order parameter: a) for a sequence of phase slips over the sample; b) amplified in a vicinity of one space-time vortex, with the corresponding contour plot. The time is measured in units of $\gamma\xi^2$.

The evolution of the phase $\vartheta$ is presented in Fig. 2. Fig. 2a shows the outer regions with the CDW almost at rest and the inner region where the local phase velocity follows closely $\omega \propto J(t)$; they are separated by steep drops of the phase $\vartheta^t$ accumulating a difference of $2\pi$ at every vortex. The vorticity density $N = (\partial_x \vartheta^t - q)/2\pi$ is shown in Fig. 2b. The density of the vorticity production $R$ in Fig. 2c show a needle-like peaks of the resolution determined width which corresponds to the expectation $\delta$-function in an ideal case.

Time and space derivatives of the order parameter phase are diverging at the vortex core. In view of the local electroneutrality Eq. (25), the phase gradient is followed by the concentration of the normal carriers which is also divergent being near zero almost elsewhere (Fig. 3a). In the expression for the electric field $E \propto \omega C^2$ the divergence of $\omega$ and the zero of $C$ compensate near the core, so only finite bumps are left (Fig. 4b) on tops of rims separating the outer region with E≈0 and the inner region with $E \propto J$.

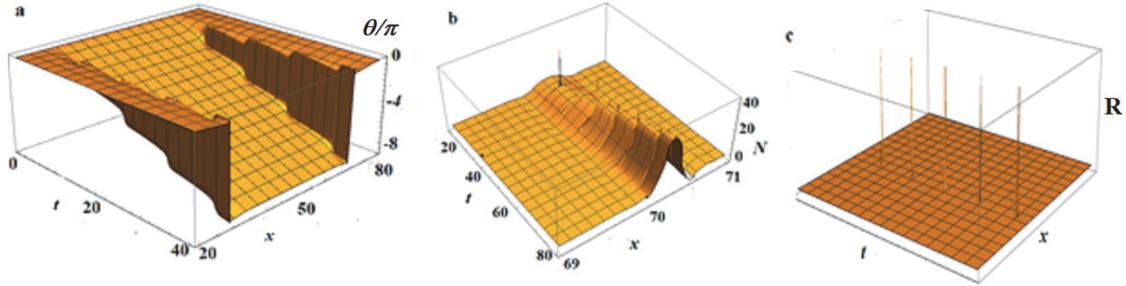

Fig. 2. Spacio-temporal evolution of the order parameter phase $\vartheta$: a) the phase, b) the vorticity density $N$, c) the density of the vorticity production $R$.

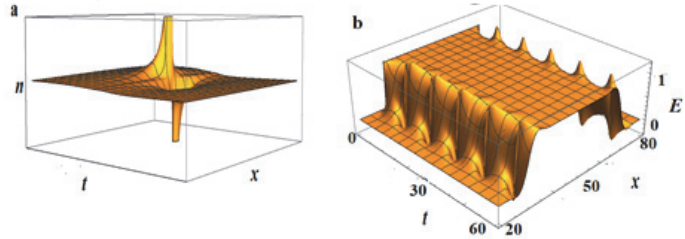

Fig. 3. 3D plots for space-time distribution of: (a) the carriers' concentration $n(x,t)$ in a vicinity of the electronic vortex following the phase gradient $q(x,t)$; (b) the electric field $E$ (normalized to the midway value $E(t,50) \approx cnst$) for a sequence of vortices.

## 7. Conclusions.

Exploiting the chiral transformation and understanding the role of the chiral anomaly allows to formulate a phenomenological theory in terms of equations for the CDW complex order parameter, the electric potential and the concentration of normal carriers. The theory resolves the problem of violation of the conservation law for condensed carriers rising dangerously for nonstationary inhomogeneous regimes; that allows to model consistently such strongly nonlinear effects as phase slips and nucleation and propagation of phase vortices.

Conceptually, the unconventional view is that the collective density and the current always correspond to the nominal number of condensed electrons as if the temperature is zero and there are no excitations above the gap. The actually present normal carriers are dragged by the phase deformations in such a way that their reaction erases the collective quantities: the normalized DOS $N_F^{-1} dn/d\zeta$ plays the role of the "normal density" $\rho_n$ eliminating gradually the "collective density" as $\rho_c = 1 - \rho_n$.

A number of approximations can be applied separately or in conjunctions: diffusion approximation for normal carriers (throughout this article), infinite normal conductivity, local electroneutrality, small gap limit. The last two conditions have been exploited here to simplify numerical solution of equations resulting in a clear picture of spontaneous sequence of phase slips generated under the injected normal current.

There is a potential to improve and expand the presented approach in future studies. The limitation to a quasi-equilibrium treatment of normal carriers can be overcome, in principle, by generalization of diffusion equations to kinetic ones. The ensemble of normal carriers can include also the gas of microscopic solitons which are recognized as major carries in systems of weakly interacting chains at low temperatures.

**Acknowledgements**
We acknowledge the financial support of the Ministry of Education and Science of theRussian Federation in the framework of Increase Competitiveness Program of NUST MISiS (N K3-2017-033).

.